\documentclass[conference]{IEEEtran}
\IEEEoverridecommandlockouts
\usepackage{cite}
\usepackage{amsmath,amssymb,amsfonts}
\usepackage{algorithmic}
\usepackage{graphicx}
\usepackage{textcomp}
\usepackage{xcolor}
\usepackage{pax}
\usepackage{url}
\def\BibTeX{{\rm B\kern-.05em{\sc i\kern-.025em b}\kern-.08em
    T\kern-.1667em\lower.7ex\hbox{E}\kern-.125emX}}
\begin{document}

\title{ChatGPT for PLC/DCS Control Logic Generation}

\author{\IEEEauthorblockN{Heiko Koziolek\IEEEauthorrefmark{1}, 
Sten Gruener\IEEEauthorrefmark{1}, 
Virendra Ashiwal\IEEEauthorrefmark{1}}
\IEEEauthorblockA{
	\IEEEauthorrefmark{1}ABB Research, Ladenburg, Germany\\
Email: $<$firstname.lastname$>$@de.abb.com}
}

\maketitle
\IEEEpubidadjcol

\begin{abstract}
Large language models (LLMs) providing generative AI have become popular to support software engineers in creating, summarizing, optimizing, and documenting source code. It is still unknown how LLMs can support control engineers using typical control programming languages in programming tasks. Researchers have explored GitHub CoPilot or DeepMind AlphaCode for source code generation but did not yet tackle control logic programming. The contribution of this paper is an exploratory study, for which we created 100 LLM prompts in 10 representative categories to analyze control logic generation for of PLCs and DCS from natural language. We tested the prompts by generating answers with ChatGPT using the GPT-4 LLM. It generated syntactically correct IEC 61131-3 Structured Text code in many cases and demonstrated useful reasoning skills that could boost control engineer productivity. Our prompt collection is the basis for a more formal LLM benchmark to test and compare such models for control logic generation.
\end{abstract}


\section{Introduction}
A generative artificial intelligence (AI) is a special kind of AI system that can generate text, images, videos, and also program code. Large language models (LLM) based on the transformer architecture have significantly raised the capabilities of generative AI for many practically relevant tasks in recent years~\cite{Zhou2023}. Popular LLMs include BERT, LaMDA, and GPT~\cite{Zhou2023}. GitHub Copilot uses GPT-3 and has provided a 55 percent developer productivity boost in experiments~\cite{Peng2023}. DeepMind's AlphaCode achieved high ratings in programming competititon tasks~\cite{Li2022}. Control engineering for industrial automation application can involve custom control logic programming, e.g., in IEC 61131-3 Structured Text (ST). Such logic can involve special mathematical algorithms, control strategies, or optimization routines.

While LLMs have produced useful code generation results for general purpose programming languages, it is not yet well understood how they can contribute to control logic engineering. Publicly available LLMs are not specifically trained for control logic programming, so it is unknown if they can generate valid source code in IEC 61131-3 ST. There is no systematic collection of domain-specific prompts for industrial automation available. It is widely unknown how useful LLMs generated source code is in practice and how to formulate prompts to yield effective results. LLMs are continuously evolving improving their answer quality over time. Metrics are needed to evaluate the quality of LLM source code generation, so that different LLMs could be compared or the effectiveness of a custom training could be verified.

Reseachers have tackled control logic generation in the last 20 years with different approaches~\cite{Koziolek2020b}. Several authors formulated control logic using UML or SysML notations, which were then translated into IEC 61131-3 ST~\cite{Vogel-Heuser2005,Haestbacka2011,Thramboulidis2011,Julius2017}. Others derived object-oriented models from piping-and-instrumentation diagrams (P\&IDs) and then applied pre-specified rules to automatically identify topological patterns and generate IEC 61131-3 ST\cite{Drath2006,Steinegger2012,Koltun2018,Koziolek2020a}. None of these approaches has gained widepsread adoption in practice, and none of these approaches utilized LLMs~\cite{Koziolek2020b}. Experiments on code generation involving LLMs largely come from the IT world and focused on Java, Python or C\# code (e.g., ~\cite{Nguyen2022,Li2022,Peng2023,Borji2023}). A detailed analysis of LLMs' capabilities to generate control logic is missing.

The contribution of this paper is a collection of 100 prompts that can be used to test an LLMs' ability to generate correct and useful control logic for industrial automation. These prompts are the result of an exploratory study. The prompt collection is i) comprehensive and systematically structured, covering different aspects of control engineering, ii) independent of any particular LLM, and iii) publicly available for independent testing and refinement by other researchers. The prompt collection is a pre-cursor for a more formal LLM benchmark for quality assessment and comparison of future LLMs specifically for control logic generation.

To assess the usefulness of our prompt collection and explore the current state of LLM control logic generation capabilities, we have fed the prompts to OpenAI's popular ChatGPT with the GPT-4 LLM and analyzed the generated answers. We found that ChatGPT can often generate sophisticated, syntactically correct IEC 61131-3 code given natural language prompts. In addition, we perceived interesting reasoning skills and a vast amount of domain knowledge retrievable from GPT-4. While the current LLMs still have limitations and occasionally generate false answer, we hypothesize that such models could boost control engineering productivitiy significantly, at least for the use case of implementing custom control logic.

The remainder of this paper is structured as follows: Section 2 surveys related work before Section 3 explains our methodology. Section 4 provides an overview of our prompt collection and Section 5 analyzes the answers obtained from ChatGPT. Section 6 discusses our findings. Section 7 lists threats to the internal, construct, and external validity of our study. Section 8 concludes the paper and formulates three hypothesis as well as future work.

\section{Related Work} 
Since the development of safe and efficient control logic is expensive, both researchers and practitioners have developed methods to generate control logic out of various artifacts in the last 20 years~\cite{Koziolek2020b}. Several rule-based methods created knowledge bases to generate control logic out of formal P\&IDs (e.g.,~\cite{Drath2006,Steinegger2012,Koltun2018,Koziolek2020a}) but were still limited by missing standard formats (e.g., ISO 15926/DEXPI). Other methods followed a model-driven paradigm and generated control logic from UML or SysML models (e.g.,~\cite{Vogel-Heuser2005,Haestbacka2011,Thramboulidis2011,Julius2017}). 

Practitioners, for example, generated control logic code skeletons by importing information from table-based I/O lists into programming environments and then adding ``glue logic'' manually~\cite{Koziolek2021}. Others encoded parts of control logic in system control diagrams (SCD), for a direct translation into IEC 61131-3 notations~\cite{Drath2018}. In addition, machine learning has been applied to aid the migration of control logic to novel notations~\cite{Mohan2019}.

Outside of industrial automation, code generation has gained significant attention with the evolution of LLMs~\cite{Zhou2023}. Due to the increasing capabilities of LLMs, few authors (e.g.,~\cite{Welsh2022}) have already speculated about ``the end of programming", meaning that the encoding of low-level algorithms will lose its importance in practice and education. For example, DeepMind's specially trained AlphaCode achieved high ratings when solving tasks from official programming competitions~\cite{Li2022}. GitHub Copilot has become popular to generate boilerplate code or to optimize existing code~\cite{Nguyen2022}. However, none of these approaches was analyzed for its capabilities regarding typical control engineering tasks.

GPT-3 was introduced in 2020 with a then-unprecedented size of 175 billion parameters. Khan and Uddin~\cite{Khan2022} demonstrated automatic code documentation generation using GPT-3 and it was deemed useful to support developers. 
The release of ChatGPT in November 2022 led to wide-spread use of Generative AI to solve text generation tasks. Borji~\cite{Borji2023} collected categories of ChatGPT failures, e.g., regarding spatial, temporal, physical reasoning, regarding logic and mathematics, and regarding coding, where he found several syntactical and semantical problems. With the release of GPT-4 in March 2023, ChatGPT can now also use this model, which according to OpenAI scores significantly higher on many LLM benchmarks. 

For prompt collection and engineering, White et al.~\cite{White2023} have compiled a number of prompt engineering patterns that can be used to improve both prompts and answers of LLMs, such as ChatGPT. In a separate work, they extended the pattern catalog specifically for improving code quality, refactoring, and software design~\cite{White2023a}. None of the existing works has specifically focused on analyzing the capabilities of LLMs to generate specifically control logic used in industrial automation. Our study is meant to explore this perspective in a first step.

\section{Methodology}
The guiding \textbf{research question} of our exploratory study was: how can LLMs support control engineers in creating control logic for industrial automation applications? To answer the question, we decided to create a representative collection of 100 prompts in 10 different categories that cover multiple aspects of control logic engineering (e.g., PLC programming tasks, sequential control logic, or interlocks). We acknowledge that custom control logic programming is rare in several application domains (e.g., rolling mills), where most logic engineering involves wiring pre-specified and often proprietary function blocks to input and output signals with a low amount of custom algorithms. While this limits the scope of control logic generation, there are still several contexts where it could improve productivity significantly.

The prompt collection aims to be reusable across different LLMs and shall be considered as a precursor for a more formal \textbf{benchmark} to quantitatively compare such LLMs. The prompt collection can be evolved and refined in subsequent research, as the understanding of LLMs increases. We fed all prompts into an LLM and analyzed the answers for plausibility and correct functionality. We provide a subjective scoring of the answer quality, since the quality is largely influenced by the quality of the input prompts. A few answers (e.g., for model-predictive control) still require more in-depths analyses. For selected answers, we entered the generated code into a PLC programming environment and checked successful compilation. %Future work could create more formal test cases to rank the outputs and provide a more refined scoring of their quality.

As \textbf{subject} for our study, we chose ChatGPT connected to the GPT-4 language model, because of its wide-spread availability and good usability. We did not use Deepmind AlphaCode, since it is trained for Java/Python/C\# code, and we did not use GitHub Copilot, since the available version used  a previous version of GPT and restricts the generation to source code, which is too limiting for control engineering. To avoid vendor bias and increase external validity, we chose to generate mostly IEC 61131-3 Structured Text (ST) programs, as one of the most popular textual control programming languages supported by many PLC and DCS development environments. ST could potentially be translated to other textual notations. Function block diagrams, ladder logic, and sequential function charts require graphical notations which are currently not directly supported by LLMs. For advanced process control prompts, we chose typical MPC languages, such as MATLAB, C++, or Python. 

We imitated typical \textbf{programming tasks} of control engineers and iteratively refined our prompts until we reached useful or interesting results. We mainly focused on control logic generation but also included a few prompts involving code understanding, bug fixing, or documentation. These could be refined and extended in future work. For several prompts (e.g., PLC programming tasks or standard algorithms) we also reference the prompt source. However, many of the prompts originate from the authors based on domain knowledge. We used a set of control narratives from large-scale automation projects carried out by ABB as inspiration, but did not use them verbatim to avoid IP issues. We also analyzed the control logic of several large ABB DCS projects in the last few years to inform some of the prompts. It is up to practitioners and researchers to judge whether the prompts are sufficiently representative, typical, and informative.

For \textbf{transparency} all prompts and answers are available as markdown, PNG, and PDF files published on GitHub\footnote{\url{https://github.com/hkoziolek/control-logic-generation-prompts}} according to the FAIR data principles\footnote{\url{https://www.nature.com/articles/sdata201618}}. We encourage other researchers to enhance and refine the prompts. Independent exact reproduction of the answers is however difficult due to the stochastic nature of LLMs.

\section{Prompt Collection}
Table~\ref {tab:promptset} provides an overview of our prompt collection. As many prompts have a significant length (e.g., including detailed domain-specific instructions), we do not show their full extent in the table, but instead use short names here and refer the reader to the full table available online. The prompt collection consist of 10 categories with 10 prompts each to cover a wide range of control logic engineering tasks. The rationale for the categories is as follows.

Prompts in \textbf{Standard Algorithms} are meant to test syntactical correctness and are inspired by typical function block libraries, such as the OSCAT Basic library\footnote{\url{http://oscat.de/}}. These prompts have a low direct usefulness, since control engineers can simply reuse the available function blocks instead of re-creating them with generative AI. They rather test an LLM's ability to comply with the syntax and constraints of the target language and the corresponding answers are comparably easy to assess.

Custom control logic in practice often comprises special \textbf{Mathematical Functions}, so the second category is meant to test the mathematical capabilities of an LLM. The prompts cover typical mathematical functions, which are meant as approximations of custom control logic. The prompts in \textbf{PLC Programming Tasks} are inspired by typical exercises in introductory PLC programming courses. They often require dealing with concrete real-world concepts. These prompts thus can test some of the spatial and physical reasoning capabilities of an LLM. In addition, these prompts could be used as templates to solve typical PLC programming problems faster using generative AI.

\textbf{Process Control} prompts involve different kinds of analog control, often involving PI or PID controllers. They cover different process control aspects, such as feedforward, cascade and ratio control. Prompts in the category \textbf{Sequential Control} test an LLM's ability to generate code for startup/shutdown sequences and batch applications, which involve an number of different steps. The correct sequence of steps for an abstractly formulated task requires specific application domain knowledge encoded in LLMs.

\textbf{Interlocks} are safety mechanisms, often linking an individual sensor reading with a concrete actuator response. While the code to express interlocks is simple, the prompts in this category also check an LLMs ability to create alternative notations (e.g., cause and effect matrices) as well as the ability to suggest required interlocks for a given situation. Prompts for \textbf{Diagnostics and Communication} test whether an LLM can deal with specific requests regarding communication protocols often linked to control logic.

The category \textbf{Advanced Process Control} includes prompts that for example ask for model-predictive control schemes in MATLAB, Python or C++, since these are typical languages for this kind of control logic. The last two categories \textbf{Various Engineering Inputs} and \textbf{Programmer Support} do not directly aim at control logic code generation, but explore different prompts that can inform subsequent code generation. They include generating control narratives, I/O Lists, or P\&IDs. Furthermore, they provide simple tests for an LLM's ability to find errors in source code provided as prompt, to explain source code, or to generate documentation.

The prompt collection may be refined and improved in future work, as a better understanding of generative AI is achieved by practitioners and researchers. Individual prompts could be enhanced with more context or completely replaced to test more specific parts of an LLM. We consider the prompt collection as a precursor for defining a more formal LLM benchmark for control logic generation.

\begin{table}[htbp]
\center
  \includegraphics[width=\columnwidth]{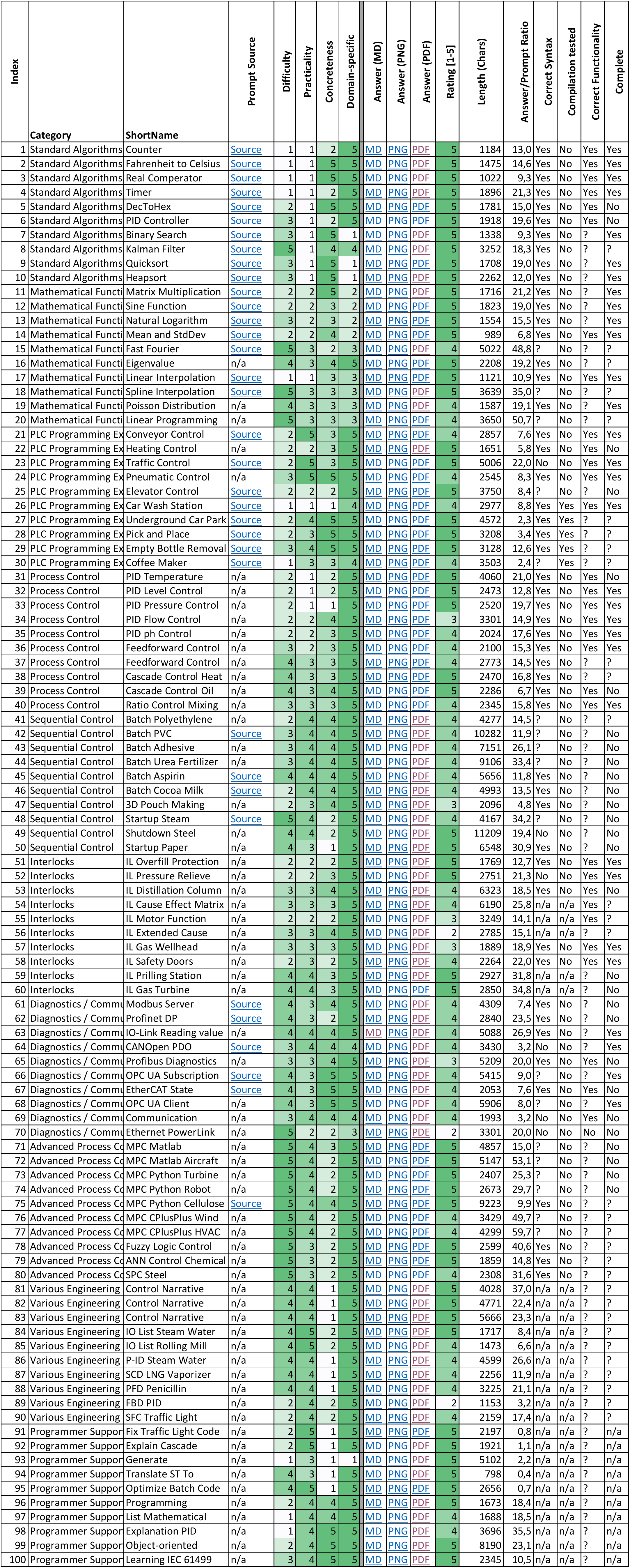}
  \caption{Prompt set with prompt scoring and answer scoring: 100 prompts for control logic generation in representative 10 categories} 
\label{tab:promptset}
\end{table}

\section{Data Collection from ChatGPT}
In the following we summarize and analyze several of the 100 answers obtained from ChatGPT using the prompts listed with short names in Table~\ref{tab:promptset}. Due the page limit we can only illustrate a few selected ChatGPT answers and refer the reader to the online resources with all answers for deeper analysis.

\begin{figure}[htbp]
\center
  \includegraphics[width=\columnwidth]{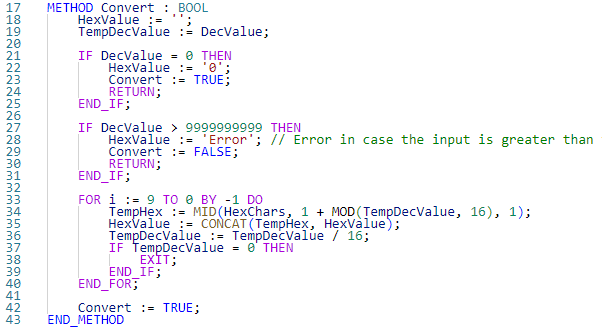}
  \caption{ChatGPT-generated IEC 61131-3 ST code for converting a decimal value to a hexadecimal string (`DecToHex', excerpt)} 
\label{fig:dec2hex}
\end{figure}

\textbf{Standard Algorithms:} 
ChatGPT generated syntactically correct code for all prompts in this category, often with useful code comments and additional explanations below the code. The generated code utilizes keywords and standard functions defined in the latest version 3.0 of the IEC 61131-3 specification, such as `METHOD' or `CONCAT'. Figure.~\ref{fig:dec2hex} illustrates an excerpt from one example answer for (`\textsc{Dec2Hex}'), a prompt asking for a function block to convert a 10-digit decimal value to hexadecimal. Besides syntactical correctness, the code is also efficient (e.g., short cutting for decimal value `0', line 21-24) and defensive (e.g., checking the upper limit of the input value, line 27). Standard functions defined in the specification are used (e.g., MID, MOD, CONCAT). Algorithms like (`\textsc{Quicksort}') in this category are not typical for control programming, yet ChatGPT generated correct variants in ST, thus hinting at the ability to handle yet unknown custom control logic. The average answer/prompt ratio is comparably high in this category (16.8), since the prompts for well-known, named algorithms are short. However the code generated in this category is usually already available from libraries.

\textbf{Mathematical Functions:}
In this category, ChatGPT demonstrated knowledge of many mathematical algorithms. For example, it provided an efficient algorithm for (`\textsc{Matrix Multiplication}') and used the standard Taylor series expansion for computing (`\textsc{Sine Function}') without explicitly being prompted for it. For (`\textsc{Fast Fourier Transform}') it generated a simplified version of the Cooley-Tukey FFT algorithm. It also used a data type 'COMPLEX' and an operator `$<<$', which are not according to IEC 61131-3 V3.0. However, after we pointed out these issues in the next prompt, ChatGPT was able to quickly generate valid code in a follow-up answer. Due to the algorithm's length, ChatGPT stopped the answer output in the middle of the generation several times, presumably because a token limit was reached. The generation could however be resumed where it stopped by simply prompting `continue'. For computing (`\textsc{Eigenvalue}'), ChatGPT reported the constraints of IEC 61131-3 regarding matrix multiplications and stated: ``[...]  I can provide you with a function block that leverages the power iteration method [...]. The power iteration method is simple and easy to implement but has limitations. It only finds the largest eigenvalue, and its convergence can be slow for certain matrices. It also doesn't guarantee convergence for all matrices."

\begin{figure}[tbp]
\center
  \includegraphics[width=0.8\columnwidth]{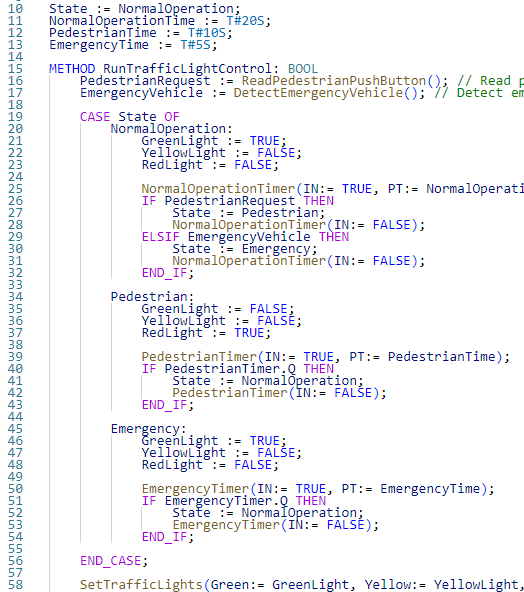}
  \caption{ChatGPT-generated IEC 61131-3 ST code for controlling a traffic light (`Traffic Control', excerpt)} 
\label{fig:tl}
\end{figure}

\textbf{PLC Programming Tasks:}
In this category, ChatGPT demonstrated specific reasoning skills on real-world concepts. As an example, Figure~\ref{fig:tl} shows an excerpt of the generated code for (`\textsc{Traffic Control}'). ChatGPT came up with different light colors, despite not being prompted for them. It also implemented a custom state machine for the phases requested in the prompt: normal, pedestrian, and emergency. It generated timers with concrete waiting times which are plausible. As the prompt was unspecific, the generated code never turns on the yellow light. Furthermore, the code requires additional code for reading the inputs and writing the outputs, which is however again a consequence of the unspecific prompt. For another example, (`\textsc{Elevator Control}'), ChatGPT broke down the code into numerous methods with concrete values. It explained that it did not yet implement an optimized scheduling for multiple concurrent requests or handling emergency situations. The generated code can be refined and extended in follow-up prompts. The prompts in this category could also be refined for specific vendor implementations and concrete input and output signal references. 

\textbf{Process Control:}
In this category, we tested generating code for different kinds of PID controllers, e.g., for level, flow or pressure control in specific situations. ChatGPT generated the according source code and also sometimes provided starting values for proportional, integral, and derivative gain tailored for the use case at hand. Besides the generated programs, it also provided explanatory text at the bottom. Several of the answers contained explicit main program loops with a sleep timer, which are not needed since the programs are assigned cyclically executing tasks in the PLC runtime. We also successfully generated a few examples for cascade control and ratio control. ChatGPT could potentially support in loop tuning by generating simulated data and providing visualizations, which is future work.

\textbf{Sequential Control:}
For sequential control, we mainly tested the generation of plausible code sequences for batch, startup, and shutdown processes. For example, the answer for (`\textsc{Batch Polyethylene}') contains a code skeleton for seven steps of polyethylene production, each with generated timers. Such skeletons could be extended manually by implementing individual steps or by generating the code for the individual step with ChatGPT. This demonstrates that also larger programs can be generated in a top-down fashion. ChatGPT can also be used to retrieve domain knowledge on specific recipes (e.g., (`\textsc{Batch Cocoa Milk}') or machines (e.g., (`\textsc{Startup Paper Machine}'). With (`\textsc{Startup Steam Generator}') we also experimented with model-predictive control for an optimized start-up process.

\begin{figure}[tbp]
\center
  \includegraphics[width=0.7\columnwidth]{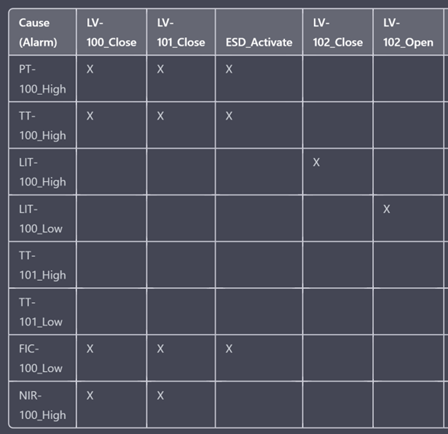}
  \caption{ChatGPT-generated IEC 62881 Cause \& Effect Matrix (excerpt). Both the tagnames and the actual interlocks were generated.} 
\label{fig:cae}
\end{figure}

\textbf{Interlocks:}
As the ST code for interlocks is comparably simple, the focus in this category was also on finding ways to elicit required interlocks in a given situation. We found for example that it is possible to build up context in a ChatGPT conversation by first prompting to generate a P\&ID in textual notation for a given situation (`\textsc{IL Distillation Column}'). The answer then provides reference points for specific sensors and actuators, which can then be used in follow-up prompts to generate required interlocking code. We also prompted for generating an exhaustive list of interlocks required in a specific situation (`IL Prilling Station'), which could support control engineers in requirements elicitation before writing code. We also experimented with alternative notations and applied the `Visualization Generator' prompt engineering pattern to instruct ChatGPT to generate cause \& effect (C\&E) matrices for interlocks according to IEC 62881. Figure~\ref{fig:cae} shows a generated C\&E matrix for a chemical reactor with plausible interlocks synthesized by ChatGPT. Such matrices are easy to review by humans and can be fed to existing ST code generators.

\begin{figure}[htbp]
\center
  \includegraphics[width=\columnwidth]{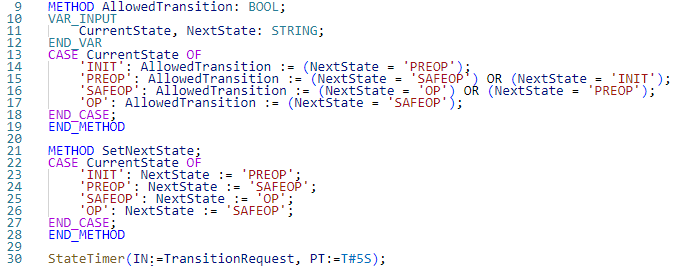}
  \caption{ChatGPT-generated IEC 61131-3 ST code for an EtherCAT state machine with correctly synthesized state transitions.} 
\label{fig:ethercat}
\end{figure}

\textbf{Diagnostics/Communication:}
In this category, ChatGPT sometimes provided very specific technical information. For example, Fig.~\ref{fig:ethercat} shows an code excerpt for (`\textsc{EtherCAT State Machine}'), which prompted for a function block for an EtherCAT slave device. The device is controlled with a state machine, where specific function are only available if the device is in a particular state. The states are predefined and only specific transitions are allowed. ChatGPT managed to encode the correctly allowed transitions in the generate IEC 61131-3 ST code, without having specific information about them in the prompt or a former conversation. An optional state `bootstrap' is missing in the code, as well as two possible transitions from OP to PREOP and INIT. However, this demonstrated knowledge of specific technical aspects of a communication protocol. In another example, (`\textsc{OPC UA Subscription Create}'), we created a function block written in C, which ChatGPT then explained how to integrate into IEC 61131-3 ST code. This demonstrates another interesting use case for code generation involving multiple languages.

\textbf{Advanced Process Control:}
In this category, ChatGPT for example demonstrated the ability to generate complex dynamic models and code for model-predictive control. For (`\textsc{MPC MATLAB Distillation Column}'), it generated a model for the dynamics of a distillation column and included a number of concrete and plausible process parameters. It used an MATLAB solver for ordinary differential equations (ODE). In a follow-up prompt, we then asked for a model-predictive control scheme for this model, which was provided by ChatGPT in a basic implementation using the `fmincon' function. We successfully executed the generated code using MATLAB. Feedback from three MPC specialists on the generated code was positive, however, a more through investigation of such code is necessary. This category also contained ChatGPT answers for fuzzy logic control, artificial neural network control and statistical process control, which were requested in Python, C++ or MATLAB.

\begin{figure}[tbp]
\center
  \includegraphics[width=\columnwidth]{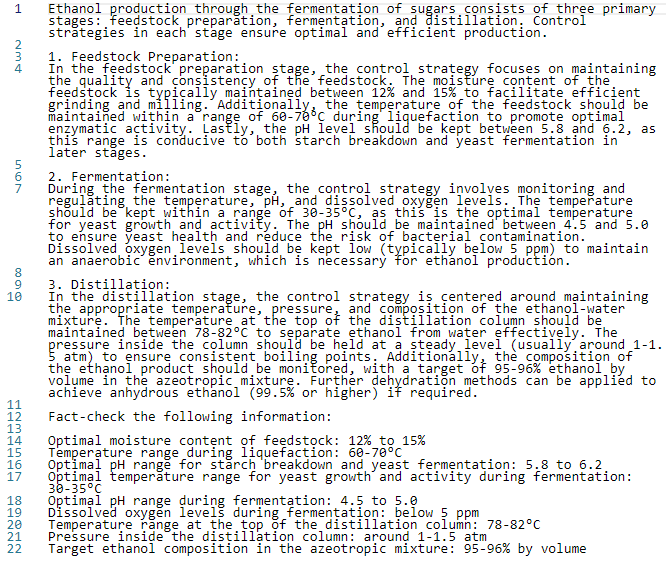}
  \caption{ChatGPT-generated control narrative for ethanol production. Generated ranges and setpoints could fact checked with other sources.} 
\label{fig:narrative}
\end{figure}

\textbf{Various Engineering Inputs:}
ChatGPT showed an ability to complete control logic requirements, by synthesizing plausible control narratives, I/O lists, or P\&IDs. For illustrative purposes, Figure~\ref{fig:narrative} shows an excerpt from a control narrative generated specifically for ethanol production. We asked ChatGPT to provide concrete setpoints and ranges, which are specially relevant for the control logic engineering. As false or hallucinated information could be harmful in this situation, we applied the `Fact Check List' pattern and asked to provide a list of facts to check at the end of the answer. This list can then be cross-verified with other sources. Mock-up I/O lists with many parameters could also be generated by ChatGPT. This ability would need to be used with appropriate contextual information in a given situation. Generating graphical notations, such as System Control Diagrams, Process Flow Diagrams, Function Block Diagrams, or Sequential Function Charts, in textual notations or ASCII art largely failed, as the ChatGPT answers were hardly useful. Future work could check to generate scripts for creating such notations indirectly using specific data formats and visualization tools.

\textbf{Programmer Support:}
Code generation can also support control engineers not only in creating code, but also in fixing bugs, optimizing algorithms, or translating existing programs. For (`\textsc{Fix Traffic Light Code}') we asked ChatGPT to find and fix errors in code generated with GPT-3.5. It correctly found four issues in the code, explained them, and then generated the revised source code. These capabilities can be used in maintaining legacy code or in refactoring. ChatGPT can also be prompted to optimize code (`\textsc{Optimize Batch Code}') or to document or explain existing code (`\textsc{Generate Documentation Urea}'). A translation between different programming languages is also conceivable, as demonstrated by (`\textsc{Translate ST to Instruction List}'). ChatGPT can also serve as a programming reference, e.g., providing keywords of the ST grammar, or list and explain the function blocks in a library. Prompts in this category potentially have high practicality and can be adapted to specific situations.

\section{Findings on GPT-4}
We analyze and interpret several of our data collection findings in the following:

\textbf{ChatGPT can generate syntactically correct ST:} This is unexpected, since according to our knowledge ChatGPT was not explicitly being trained for it. OpenAI has not provided the concrete training data for GPT-4, but it can be assumed that it contained a sufficient amount of ST code (e.g., from a GitHub snapshot) or the specification itself. ChatGPT uses the latest language features (e.g., OO-programming) and standard functions defined in the specification, which may not be supported by all PLC environments. However, it is conceivable to instruct ChatGPT to adapt the generated code to the limitations of a particular PLC environment and for example generate `FUNCTION\_BLOCK' instead of `METHOD'. Many of our generated code samples appeared efficient and robust, thus directly useful in practice. ChatGPT is also aware of IEC 61131-3's limitations, and for example does not try to generate low-level code for tasks unsuited for the language.

\textbf{ChatGPT can generate large programs top-down:} ChatGPT-generated code is usually limited to around one or two screens of text. However, the answer for a larger problem can be broken down into several separate steps for which then individual code can be generated in follow-up prompts, as seen in our sequential control prompts. This potentially allows to generate also larger programs, beyond ChatGPT's token limits. In general, useful code generation often originates from a longer conversation with several prompts, where the initially generated code is subsequently enhanced, refined, and optimized. 

\textbf{ChatGPT can provide domain knowledge:} control engineers sometimes struggle with incomplete requirements, which can require time-consuming and error-prone feedback loops with process engineers. The training set of GPT-4 contains vast amounts of production domain knowledge, which can be retrieved with the right prompts. ChatGPT can then provide concise answers to specific questions, not requiring the control engineer to read long instructional texts or production handbooks. ChatGPT has physical and common sense reasoning skills which can speed up solving real-world automation problems. Its statistics can simulate an \textit{understanding} of a traffic control light or a turbine in a power plant, even though ChatGPT does not include an explicitly encoded `world model'. Neural networks in GPT-4 can synthesize useful text patterns based on statistics around these situations. Besides production domain knowledge, ChatGPT also indicated deep technical knowledge about communication protocols, or specifics of the ST language. The exact limits of this knowledge still need to be understood in more detail (e.g., non-public, vendor-specific data is not accessible for GPT-4) but we found that ChatGPT can already tackle many situations without custom training.

\textbf{Prompt fidelity largely determines answer fidelity:} in our experiments, one of the most crucial factor for code generation quality was the prompt granularity. Unspecific prompts lead to generic, high-level code fragments, while concrete, specific prompts can lead to very detailed programs. Asking the right question in a given situation is however challenging. ChatGPT can be used for prompt engineering, successively refining prompts using patterns, such as \textit{Question Refinement}, \textit{Alternative Approaches}, \textit{Cognitive Verifier}, or \textit{Refusal Breaker}. However, especially regarding the streamlined syntax of ST this has limits. In many cases it can be faster for an experienced control engineer to directly formulate the control logic in ST, which is a much more condensed notation than the natural language required for ChatGPT prompts. ST code often depends on many specific signal references, which are wired together using sophisticated black box function blocks from vendor libraries. For these situations, using ChatGPT code generation and formulating many natural language prompts may require more efforts than simply writing the code manually.

\textbf{ChatGPT-generated code can be significantly improved by more context:} we found that it is often advisable to build-up sufficient context in a number of prompts, before prompting for an actual code generation. ChatGPT can be instructed to generate signal lists or textual P\&IDs, capturing information required for code generation. This simulates the situation of control engineers processing textual control narratives and checking the corresponding graphical P\&IDs. Textual P\&IDs are so far still vastly inferior to the information encoded in real P\&IDs, but it is conceivable that future GPT-versions could work with standardized P\&ID data models, such as ISO 15926/DEXPI, as input and then base the code generation on these. %For MPC, first a dynamic model of an equipment can be generated, before synthesizing the actual MPC code. 

\textbf{ChatGPT answers require careful validation:} many times, ChatGPT answers appear correct and well-formulated but can also contain subtle errors or omissions. Syntactical problems can be easily identified by attempting to compile the generated code. Semantic problems may be harder to find. Testing the control code without physical I/O may require simulated input signal values, which could also be synthesized with ChatGPT. However, ChatGPT answers often state that generated code is meant as an example and not for direct use in production. Oversight of experienced control engineers is still required. When generating control narratives, we applied the \textit{Fact Check} pattern, so that ChatGPT also provides a list of facts to cross-check with other sources.

\textbf{GPT-4 limitations:} Typical control narratives contain dozens of pages, which are still beyond GPT-4's input token limit, so it is not yet possible to enter non-trivial control narratives for a large generation step. Control narratives are often accompanied by graphical data, whose processing with ChatGPT we could not evaluate in the currently available version that only accepts textual inputs. We also noticed for many complex programs that ChatGPT stopped the code generation due to a token limit and then required prompting 'continue'. These issues may be resolved in future versions.

\section{Threats to Validity}
Our exploratory study targets exploring the capability of Generative AI to generate control logic. It is affected by the following threats to validity: 

\textbf{Internal validity} refers to the extent in which a study's design and execution establish a causal relationship between independent and dependent variables. Independent variables are our prompts, which can be tested by other researchers against different LLMs. The dependent variables are the quality measures on the ChatGPT answers. We found a causal relationship between more concrete and specific prompts leading to higher-quality answers and provided many examples for these. For transparency our obtained results are fully published online for independent re-assessment.

While ChatGPT and other LLMs do not produce outputs deterministically, our results are comparably easy to reproduce (albeit not syntactically) for third parties to increase trust in the internal validity. Our prompt collection accounted for breadth and depths, but may not yet be fully representative for all kinds of control logic or text-related control engineer tasks. Many prompts originated from the authors and are not triangulated with other sources. Another threat to the internal validity is the still informal quality scoring of the ChatGPT outputs, as we have not yet defined an objective metric for the answer quality. Many of the answers have not yet been compiled and thoroughly checked for functional correctness.

\textbf{Construct validity} evaluates whether a study accurately measures the intended theoretical constructs. We applied the prompts to ChatGPT and GPT-4, which can be considered a very popular LLM, thus an appropriate substitute for generative AI. We generated IEC 61131-3 ST as a representative textual control programming language, which is standardized and supported by many PLC and DCS programming environments. Ladder diagrams are more popular in Nothern America, but difficult to generate due to their graphical nature. It is conceivable that generated ST code could be transformed to ladder diagrams. We encoded typical programming exercises and common function blocks in our prompts, which should support their construct validity for typical tasks. 

\textbf{External validity} gauges the generalizability of a study's findings to other populations, contexts, or times. Our prompts are not ChatGPT-specific and can be reused on other LLMs that support natural language inputs. The generated code can be compiled by standard-compliant IEC 61131-3 programming environments. Vendor-specific flavors would need to be accounted for, but according variations of the answers could be quickly generated with refined prompts. Since our prompts reflect typical function blocks and basic algorithms, generating ``similar'' kind of code with slightly adapted prompts should be possible. This could for example be used to support implementation of custom control logic with specific mathematical functions.

\section{Conclusions and Future Work}
The following hypotheses could be tested in follow-up research, e.g., in controlled experiments:

\begin{itemize}
\item Hypothesis 1: In the context of custom control logic implementation using textual programming languages, LLMs can significantly increase control engineer productivity for typical programming tasks.
\item Hypothesis 2: LLMs are significantly more \textit{efficient} for eliciting specific, established automation engineering domain knowledge than Internet search engines (e.g., Google), typical question and answer websites (e.g., stackoverflow), social media sites (e.g., Reddit), or monographs (e.g,. production handbooks).
\item Hypothesis 3: Control engineers well-trained in prompt engineering can achieve significantly higher productivity for solving typical custom control logic programming tasks using LLMs than untrained control engineers.
\end{itemize}

Reseachers and practitioners can directly use the 100 prompts created for this work and adapt them to specific contexts or test the output of different LLMs. The prompt collection is a starting point to be refined and extended in future versions. A formal quality scoring method for LLM-generated answers could turn the prompt collection into a benchmark for assessing and comparing LLMs. Inspired by generic prompt engineering patterns, it is conceivable to create special prompt engineering patterns for control programming. Future work could also tackle a quantification of the productivity and efficiency improvements in case studies and controlled experiments.

\bibliographystyle{IEEEtran}
\bibliography{etfa2023}

\end{document}